# Event-Driven Simulation for Rapid Iterative Development of Distributed Space Flight Software

Toby Bell
Stanford University
496 Lomita Mall, Stanford, CA 94305
tbell@cs.stanford.edu

Simone D'Amico
Stanford University
496 Lomita Mall, Stanford, CA 94305
damicos@stanford.edu

*Abstract*— This paper presents the design, development, and application of a novel space simulation environment for rapidly prototyping and testing flight software for distributed space systems. The environment combines the flexibility, determinism, and observability of software-only simulation with a level of fidelity and depth normally attained by real-time hardware-in-the-loop testing. Ultimately, this work enables an engineering process in which flight software is continuously improved and delivered in its final, flight-ready form, and which reduces the cost of design changes and software revisions with respect to a traditional linear development process. Three key methods not found in existing tools enable this environment's novel capabilities: first, a hybrid event-driven simulation architecture that combines continuous-time and discrete-event simulation paradigms; second, a design for lightweight application-layer software virtualization that allows executing compiled flight software binaries while modeling process scheduling, input/output, and memory use; and third, high-fidelity models for the multi-spacecraft space environment, including for wireless communication, relative sensing such as differential GPS and cameras, and flight computer system health metrics like heap exhaustion and fragmentation. The simulation environment's capabilities are applied to the iterative development and testing of two flight-ready software packages: the guidance, navigation, and control software for the VISORS mission, and the Stanford Space Rendezvous Laboratory's software kit for rendezvous and proximity operations. Results from 33 months of flight software development demonstrate the use of this simulation environment to rapidly and reliably identify and resolve defects, characterize navigation and control performance, and scrutinize implementation details like memory allocation and inter-spacecraft network protocols.

TABLE OF CONTENTS

## 1. INTRODUCTION

Distributed space systems, comprising multiple smaller spacecraft working collaboratively, offer benefits and challenges over traditional monolithic spacecraft. Their modularity allows for cost-effective scalability, enabling a range of missions from Earth observation to deep space exploration. Distributed space systems support diverse mission architectures, such as swarms or constellations, which can improve spatial and temporal coverage, enhance data collection, and enable complex operations like interferometry or synthetic aperture radar. Missions like GRACE[1], PRISMA[2], TanDEM-X, Starling[3], and MMS[4], among others, have demonstrated the power of distributed spacecraft in orbit and have driven interest in future missions such as the Virtual Super-Resolution Optics Reconfigurable Swarm (VISORS)[5] and the Space Weather Atmospheric Reconfigurable Multi-Scale Experiment (SWARM-EX)[6]. Distributed space systems also face notable challenges, particularly in coordination and communication. Synchronizing multiple spacecraft requires advanced navigation, control, and autonomy to ensure precise operation, which increases mission complexity. Additionally, reliable inter-satellite communication is often critical to the function of a distributed space system, demanding robust, low-latency networking. Despite these hurdles, distributed space systems represent a transformative approach to space missions, offering increased capabilities and the potential for groundbreaking science.

Due to the greater operational complexity of multiple spacecraft working together, advanced on-board navigation and control software often plays a heightened role in the functioning of a distributed space system. For example, state-of-the-art navigation systems like DiGiTaL[7] and ARTMS[8] have arisen specifically to meet the needs of distributed space systems, as have novel impulsive[9] and low-thrust[10] control methodologies for spacecraft relative motion. The design and implementation of such flight software algorithms can add significant complexity[11], cost, and risk[12] to the development of a space mission. At the same time, reduced launch costs have opened space to smaller state, commercial, and academic players with less capital[13], and miniaturized spacecraft have lowered budgets and shortened development times[14], creating a need to enable developing capable flight software rapidly and for low cost.

Simulation is a critical technology for space flight software development, allowing verifying mission requirements before deployment in space for low cost. State-of-the-art simulation tools used for commercial and academic space applications include graphical mission-planning tools STK[15], GMAT[16], and FreeFlyer; programmable simulation frameworks like MATLAB/Simulink[17], *Trick* by NASA[18], and *Basilisk* by CU Boulder AVS Lab[19]; and full-system simulators like NOS3[20], 42[21], and Wind River Simics[22]. Despite the diverse capabilities offered by these tools, simulation is usually insufficient to fully capture the behaviors of the real system and must be augmented with thorough real-time

software- and hardware-in-the-loop testing to detect and resolve additional defects.

*Linear Software Development*—It is common in the space industry to use a linear development model, sometimes called the waterfall model, for flight software. The European Space Agency uses a process with five stages, from *specification* to *acceptance* (Figure 1)[23]. Linear software development was adopted from the hardware manufacturing industry, where it was highly appropriate given design changes become prohibitively expensive as development progresses. The United States Department of Defense adopted the waterfall model as standard for software development via standard DOD-STD-2167 in 1985[24].

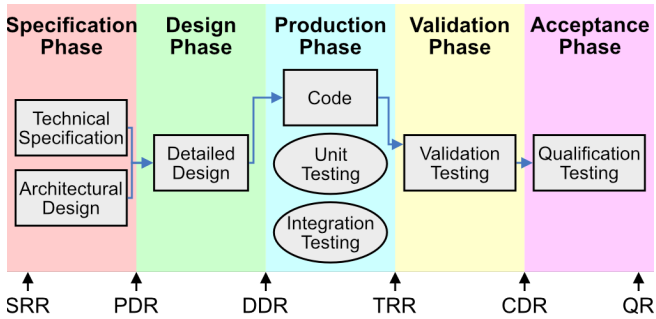

**Figure 1.** European Space Agency's linear software development model.

*Iterative Software Development*—In contrast to the space industry, the consumer software industry has shifted almost entirely to an iterative development model, and most practitioners consider frequent revisions and adaptability to changing requirements to provide greater value-for-cost than structured sequential processes[25]. In 2012 the United States Federal Aviation Administration adopted DO-178C, which provides for increased used of iterative development models for aeronautical flight software[26]. While DO-178C supplanted DOD-STD-2167, iterative software development is still comparatively uncommon in the space industry.

*Challenges of Iterative Development in Space*—Unlike consumer software, space flight software is tightly integrated with complex hardware systems onboard a spacecraft. This makes iterative development difficult, since effective integrated testing of the complete system requires access to those hardware systems for the software to behave properly. Distributed space flight software is even harder to test, requiring networked hardware for multiple spacecraft in order to test interactions between them. This dependency on a specific hardware configuration can be a significant hurdle for software teams, who now experience contention for hardware in order to test software. This hurdle can be avoided by mocking the necessary hardware in simulation: a common method for space flight software development is to prototype and simulate algorithms in a dynamic language like MATLAB or Python and use code generation to convert the tested algorithms to flight-ready C89 source code[27], for example as in the PRISMA mission[28]. This can work well, but it has downsides: the generated code may not give as much control or performance as hand-written C code, and in case of defects identified after integrating the generated code with other systems, it is not always obvious how defects in the generated code map back to defects in the prototype code, preventing efficient iterations.

*Hardware Testing*—Running on real hardware provides the highest fidelity possible for testing space flight software, and will often detect software implementation defects not caught by simulation testing, including due to memory use, runtime, networking, and process scheduling. For example, the Starling mission experienced unforeseen software issues in orbit because of memory exhaustion[29]; other missions have experienced anomalies due to unpredictable operation scheduling and execution timing[30]. However, hardware testbeds may not be available for low-cost missions, or may support testing only one instance of a spacecraft at a time. For example, during development for the VISORS mission, the main hardware testbed at Georgia Tech's Space Systems Design Lab could only model a single spacecraft[31], but the mission depended on autonomous interaction between two spacecraft. This made it impossible to test distributed GNC capabilities on the testbed hardware. The lack of availability of hardware to small and low-budget teams, as well as the relatively greater time cost of hardware testing for missions of any size, underscores the need for new simulation-based testing methods that can better identify low-level software defects while preserving efficient iteration.

The remainder of this paper presents the design and application of a novel simulation environment for rapid, iterative development of flight software for distributed space systems, which offers a fidelity closer to hardware testing while preserving flexibility and speed of a software-only environment. Sections 3, 4, and 5 present three key methods not found in existing tools that enable this environment's novel capabilities:

1. A hybrid event-driven simulation architecture that combines continuous-time and discrete-event simulation paradigms (Section 3)
2. A design for lightweight application-layer software virtualization that allows executing compiled flight software binaries while modeling process scheduling, input/output, and memory use (Section 4)
3. High-fidelity models for the multi-spacecraft space environment, including models for wireless communication, relative sensing such as differential GPS and cameras, and flight computer system health metrics like heap exhaustion and fragmentation (Section 5)

Finally, Section 6 discusses the results of applying the simulation environment to iterative development of two advanced flight software packages for distributed space systems—the guidance, navigation, and control software for the VISORS mission (VISORS GNC)[32], and the software kit for rendezvous and proximity operations (RPO Kit)[33]. Both packages are developed by the Stanford Space Rendezvous Laboratory, and implement advanced navigation and control algorithms for multi-spacecraft systems, including relative orbit determination and control, collision avoidance, inter-satellite communication, and autonomy. VISORS is a two-CubeSat distributed telescope mission set to launch in 2025 requiring autonomous, centimeter-precise alignment at 40-m inter-spacecraft separation. RPO Kit is a modular, flexible navigation and control subsystem for autonomous rendezvous with cooperative or noncooperative targets using fused GPS and vision-based sensing at high dynamic range. Ultimately, iterative development using event-driven simulation is applied to both projects and shown to enable rapid and thorough debugging and performance characterization.

## 2. PROBLEM STATEMENT

The objective of this work is to develop a novel simulation environment that enables rapid, iterative development of flight software for distributed space systems and, further, to evaluate its effectiveness when applied to develop two flight-ready navigation and control software packages for state-of-the-art multi-spacecraft missions: VISORS GNC and RPO Kit by the Stanford Space Rendezvous Laboratory.

The resulting simulation environment should offer the capabilities of dynamics-focused navigation and control simulations as might be implemented in MATLAB/Simulink while also exercising behaviors usually checked by real-time integrated testing on hardware. In particular, it should offer the determinism, speed, and observability of a prototyping environment like MATLAB while also ensuring that flight algorithms are robust to real-world imperfections like limited memory, indeterminate process scheduling, and unreliable wireless communication. It should operate directly on the final form of deliverable flight software, such as compiled C or C++ code, rather than simplified or analogous models, in order to enable continuous refinement and delivery of flight software. Crucially, for distributed space systems, the simulation should provide good support for modeling and debugging the complex autonomous interactions that arise within a communicating multi-spacecraft system.

## 3. EVENT-DRIVEN SIMULATION

The first of three core components of this work is a hybrid event-driven simulation architecture for the execution of distributed space flight software. The simulation architecture combines aspects of typical continuous-time and discrete-event simulation methods to precisely model effects on the scale of microseconds (such as software execution, communication delays, thruster firing, and variable-rate measurements), while also modeling the continuous orbit dynamics of a multi-spacecraft system on the scale of multiple orbit periods. Fully simulating the coupled effects of continuous dynamics and discrete state changes produces a flexible and efficient simulator capable of modeling both large-time-step orbit evolution in free-motion and millisecond-long thruster activations, for example. This ultimately enables simulating full execution of distributed space flight software with both accuracy and speed, making it suitable for rapid prototyping, iterative development, and rigorous performance testing.

*Continuous-Time Simulation*

In a continuous or time-driven simulation, the system state is propagated repeatedly by a fixed or variable time step, often using numerical integration of a model of the system dynamics to compute the state changes (Figure 2). Mathematically, a continuous simulation can be defined by a state space $S$, propagation function $f : S \times \mathbb{R} \to S$, time step $\Delta t \in \mathbb{R}$, and initial state $x_0 \in S$:

> **given** $x_0 \in S$
> **given** $\Delta t \in \mathbb{R}$
> **for** $i = 0...\infty$
> $\quad x_{i+1} = f(x_i, \Delta t).$

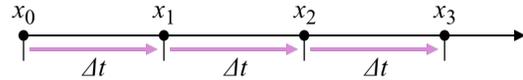

**Figure 2.** Typical time-driven simulation: successive states computed via continuous dynamics at regular intervals.

Continuous simulation is commonly used to test guidance, navigation, and control algorithms for spacecraft, since spacecraft free motion is described well by a continuous dynamical system. A weakness of time-driven simulation is its inability to natively incorporate instantaneous or discrete state changes with both speed and accuracy, even when using a variable time step: too large a step loses high-frequency effects, while too small requires too many steps for fast execution. Tools like 42, Simulink, STK, GMAT, and FreeFlyer all make use of continuous simulation, with the associated consideration for time steps or sample times.

*Discrete-Event Simulation*

Discrete-event simulation models the evolution of a system over time as a sequence of events, where the occurrence of each event produces a state change (Figure 3). A set of upcoming events is maintained, and at each step the event with the smallest time is dequeued and applied to the state (see RemoveMin below). Events may be of different types, and state changes may themselves produce additional events to be added to the event set. Mathematically, a discrete-event simulation can be defined by a state space $S$, event space $\mathbf{E}$, event application function $g : S \times \mathbf{E} \to S \times \mathscr{P}(\mathbb{R} \times \mathbf{E})$, initial state $x_0 \in S$, and initial event set $s_0 \subseteq \mathbb{R} \times \mathbf{E}$, where $\mathscr{P}$ denotes the power set:

> **given** $x_0 \in S$
> **given** $s_0 \subseteq \mathbb{R} \times \mathbf{E}$
> **for** $i = 0...\infty$
> $\quad e_{i+1}, s'_{i+1} = \text{RemoveMin}(s_i)$
> $\quad x_{i+1}, v_{i+1} = g(x_i, e_{i+1})$
> $\quad s_{i+1} = s'_{i+1} \cup v_{i+1}.$

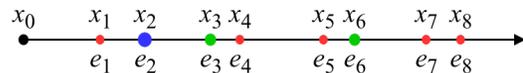

**Figure 3.** Typical discrete-event simulation: successive states computed by applying a sequence of heterogeneous events.

Discrete-event simulation often models systems where state changes are mostly instantaneous, intervals between events are irregular or uncertain, and the system state does not change significantly between events—for example, process engineering for manufacturing[34], healthcare[35], business op-erations[36], digital logic design[37], and computer networks[38][39]. A strength of discrete-event simulation is its combination of accuracy and execution speed, since time is only spent computing state updates, and long gaps without events can be skipped entirely. However, its assumption that state does not change between events makes it incompatible with many problems, including space flight.

*Hybrid Event-Driven Simulation*

Continuous and discrete-event simulation are often posed as opposite one another and specialized for contrasting purposes[40][41]. However, neither on its own is well-suited to simulate space flight software systems, which involve high-frequency discrete state changes like software state machines, timestamped measurements and actuations, and command and log messages, but also interact with a naturally continuous dynamical system in the spacecraft's orbit environment. This work develops a hybrid of the continuous and discrete-event methods in order to efficiently simulate a multi-spacecraft software execution environment. In this hybrid event-driven model, the flow of time is driven by an ordered event set as in discrete-event simulation, but continuous dynamics are also applied to the state during the variable-length intervals between events (Figure 4). A hybrid event-driven simulation can be defined by a state space $S$, event space $\mathbf{E}$, propagation function $f : S \times \mathbb{R} \rightarrow S$, event application function $g : S \times \mathbf{E} \rightarrow S \times \mathscr{P}(\mathbb{R} \times \mathbf{E})$, initial state $x_0 \in S$, and initial event set $s_0 \subseteq \mathbb{R} \times \mathbf{E}$, where $\mathscr{P}$ is the power set:

**given** $x_0 \in S$
**given** $s_0 \subseteq \mathbb{R} \times \mathbf{E}$
**init** $t_0 = 0$
**for** $i = 0...\infty$
$\quad t_{i+1}, e_{i+1}, s'_{i+1} = \text{RemoveMin}(s_i)$
$\quad \Delta t_i = t_{i+1} - t_i$
$\quad x'_{i+1} = f(x_i, \Delta t_i)$
$\quad x_{i+1}, v_{i+1} = g(x'_{i+1}, e_{i+1})$
$\quad s_{i+1} = s'_{i+1} \cup v_{i+1}.$

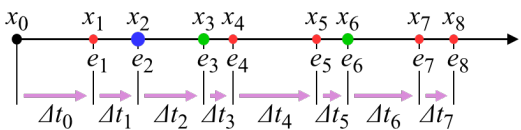

**Figure 4.** Hybrid event-driven simulation: successive states are computed by both applying events in order and modeling continuous dynamics at irregular intervals.

*Generalization*—Note, the hybrid event-driven form reduces exactly to the discrete-event form by substituting the identity propagation function $f(x, \cdot) = x$. Similarly, the hybrid event-driven form reduces exactly to the continuous-time form by substituting the identity event application function $g(x, \cdot) = x$ and initial event set $\{(i\Delta t, \cdot) \mid i = 1...\infty\}$. In this way, the hybrid event-driven simulation model can be seen to be a generalization of the continuous-time and discrete-event methods. It can also be thought of as continuous simulation with the addition of an event set, or as discrete-event simulation with the addition of continuous dynamics.

The hybrid method provides great power for the simulation of spacecraft flight software, because the explicit event set allows efficiently adjusting the time step. A simulation can quickly execute many tightly-packed events, such as impulsive maneuvers or communicating software processes, without slowing down the simulation during long periods of free motion that could be computed with a single propagation.

The notion of combining discrete-event and continuous-time simulation is not fundamentally new and has been explored previously for the purpose of analog circuit design[42], embedded software for robotics[43][44], and power grid design[45]. Combining discrete and continuous simulation is also supported by the general-purpose simulation platform Modelica[46]. However, the method appears less developed than either continuous or discrete-event simulation in literature, and its use is not widely discussed compared to continuous simulation for testing and developing spacecraft guidance, navigation, and control software.

*Implementation*

The event-driven simulation architecture in this work is implemented in C++ following a typical design for a software event-loop. The event set uses a min-heap to store pending events. Event records contain a timestamp, a function pointer, and associated data (Figure 5). The function pointer defines the event's state change, which is applied to the global simulation state simply by calling the function pointer. For example, to simulate a thruster valve opening for 1 second, one might enqueue an event at $t = 0$ with a pointer to a function that sets the valve to open, and an event at $t = 1$ s with a pointer to a function that sets the valve to closed. Events are dequeued and executed in a loop.

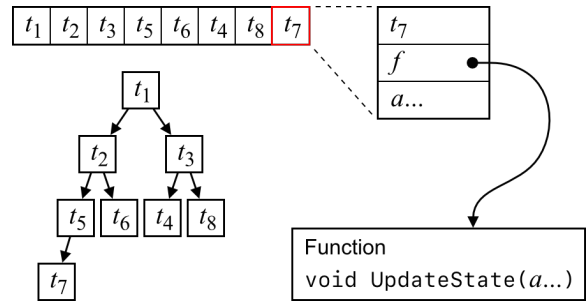

**Figure 5.** Implementation of simulation event loop as a min-heap with function pointers for events.

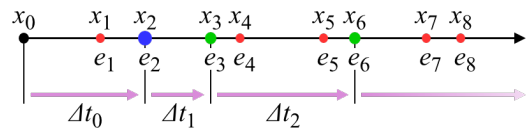

**Figure 6.** Lazy propagation of continuous dynamics for only events that need it; red events do not depend on the continuous system state.

A small optimization is made to reduce the time spent propagating the continuous system dynamics. Not all events depend on the continuous system state; some events represent purely discrete changes in other systems—for example, delayed execution of a software routine. Rather than propagate the continuous state between all events, an event must explicitly request the continuous system state via a function call when it needs it, at which point the state is lazily propagated forward if necessary (Figure 6). This allows using fewer propagations and larger time steps.

## 4. INTERFACE VIRTUALIZATION

The second primary contribution of this work is a design for interfacing with unmodified flight software in simulation. Since this work aims to enable rapid evaluation and iteration of flight-ready multi-spacecraft navigation and control software, seamlessly executing flight software in simulation is critical. To do this, multiple flight software instances must be able to interact with the simulation via a virtual interface, such that they behave identically when executed within the true multi-spacecraft system (Figure 7).

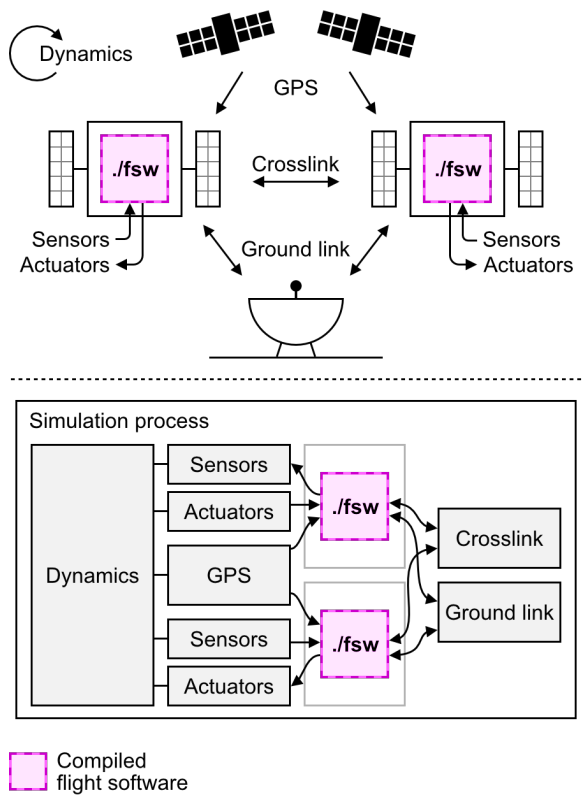

**Figure 7.** Example distributed space system running compiled flight software, and corresponding simulation environment running the same compiled flight software through a virtualized interface.

The virtual interface in this work is designed to achieve five central capabilities, summarized here and expounded in the following paragraphs:

1. Run unmodified compiled flight software
2. Run multiple interacting instances of flight software
3. Yield a deterministic total order of events
4. Run faster than real-time
5. Support easy debugging with off-the-shelf debuggers

*1. Run Unmodified Compiled Flight Software*—Some space simulation environments use software models or prototypes rather than flight-ready software, which reduces testing's effectiveness. For example, the 42 simulation environment includes representative models rather than flight-ready software[47]. For another example, a practical flight software development method is to prototype navigation and control code in MATLAB or Python, test it in simulation, and subsequently port it to C or C++ either manually or with tools like MATLAB Coder[48]. This validates the algorithm design but not the final implementation. In these cases, real-time software- or hardware-in-the-loop testing can be performed on the final implementation, but fixing defects is much harder at this stage, reducing the capacity for rapid iterative development. An effective simulation–software interface should directly run flight-ready software in order to enable rapid iterative development.

*2. Run Multiple Instances*—Simulating multiple interacting software processes is clearly needed for the development of interacting multi-spacecraft flight software. This capability can be nontrivial: until recently, the Basilisk space simulation framework did not easily support multi-spacecraft simulation by design of its message-passing framework[49]; commercial simulation products from Blue Canyon Technologies for software- and hardware-in-the-loop testing of commercial spacecraft buses only support single-spacecraft operation[50], which was a limiting factor during development of the VISORS mission[51].

*3. Deterministic Total Order of Events*—Although the execution of a distributed system in general may not be serializable[52], it's possible to design flight software to produce executions that are always serializable, for example by using message-passing between spacecraft as in the actor model[53]. Even so, it is understood that the sequence of events in a distributed system (that is, the happens-before relation) is not a total order, but only a partial order[54]. This is especially true in distributed space systems, since communication may happen over long distances between actors with large relative velocities. Thus, the total order of events, and thus the result of executing an autonomous distributed space system (such as the navigation states and control outputs produced by flight software) is inherently indeterminate. This is part of what makes creating distributed space flight software difficult and simulation critical. For the sake of analysis and debugging, it is useful for execution to be deterministic during development, so that simulations can be run repeatedly to diagnose defects. This requires a consistent total order on events. This is not possible using real-time software- or hardware-in-the-loop integration testing, in which various instances of flight software run in separate processes, since operating system schedulers and separate devices only respect a partial order per the happens-before relation. Therefore, a development simulation should be designed to select a consistent ordering of events every time it runs. Note, the total order can still be stochastic with respect to pseudorandom noise, for example for Monte Carlo testing.

*4. Run Faster than Real-Time*—For rapid iterative development, flight software should be able to be simulated faster than real-time. Low-thrust control algorithms may operate on the scale of months[55], memory leaks and fragmentation may arise slowly[56], and logical edge cases may present themselves many weeks into a mission. Tools like the open-source NOS3 and commercial real-time dynamics processor from Blue Canyon Technologies[57] support realistic integrated testing of interacting flight software components, but they run in real-time and are intended for human-in-the-loop operation, which limits their use for quickly assessing flight software performance over the course of a mission.

*5. Easy Debugging*—Debugging distributed systems is difficult and can require specialized techniques beyond those commonly used for centralized software, such as integrated testing, model checking, theorem proving, record and replay, tracing, log analysis, and visualization[58]. Meanwhile, centralized software, which usually has a well-defined total execution order (see capability (3) above), is simpler to debug with interactive debuggers[59] or even diagnostic print statements. An ideal virtual interface for software simulation should allow using print statements and off-the-shelf debuggers like GDB[60] and LLDB[61] to inspect live flight code during iterative development.

In pursuit of the five capabilities identified above, this work uses a lightweight virtual software interface in which flight software is compiled to shared libraries that implement algorithms with event-driven inputs and outputs. These shared libraries are loaded by the simulation and allowed to execute as multiple interacting virtual processes by a deterministic scheduler within a single thread. Further, `malloc`/`free` are overridden while executing flight software in order to allow simulating dynamic memory allocation for each spacecraft.

*Flight Software Shared Libraries*

The foundational design choice for the interface between simulation and flight software is the definition of the flight software deliverables as shared libraries rather than executable programs. In this design, flight software source code is compiled into shared libraries that can be loaded into either the simulation environment during development or into a host process on a real spacecraft (Figure 8). The library's interface is designed to be simple to implement in both simulation and the real flight computer environment. It's worth noting that the flight computer host process consists by definition of code that cannot be tested in simulation, so design decisions should be made to keep its implementation as simple as possible. This design supports capability (1) *run unmodified flight software*, since the shared library can be loaded in the same form by both the simulation and the real flight computer. It supports capability (2) *run multiple instances*, because, unlike executable binaries, multiple shared libraries can be loaded alongside each other and simulation code. It also supports capability (5) *easy debugging*, since all code can run together in a single process, avoiding the complexity of multi-process debugging.

Isolation between the simulation and flight code is enforced during the compilation process; while simulation code is allowed to refer to flight code, referring to simulation code from flight code triggers a compiler error.

*Event-Driven Input and Output*

The flight software shared libraries contain an initialization function for creating stateful instances of flight software, which can then be interacted with via explicit function calls to provide inputs and receive outputs. All execution of the flight software happens in response to some input. In VISORS GNC and RPO Kit, outputs are received via callback function as in the dependency injection[62] and delegation[63] design patterns, but this is an implementation detail. The simulation could be adapted to mesh with a different data output interface, such as message queues or return values. Event-driven, non-blocking flight software supports capability (4) *run faster than real-time* by only requiring execution of flight code in response to inputs that may actually cause it to take an action, rather than continuously running a 10 Hz main loop, for example. This allows the simulation to skip over longer periods of idle time in the flight software.

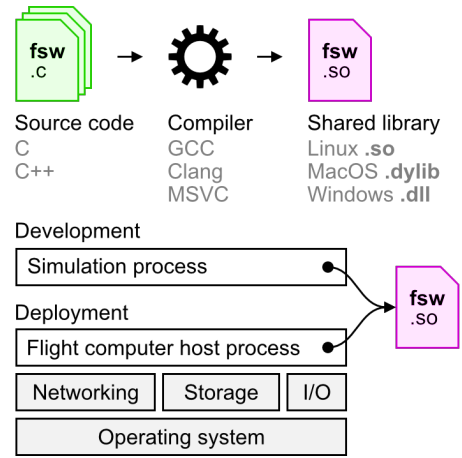

**Figure 8.** Flight software compiled as a shared library rather than executable, to load into simulation processes or deployed flight software processes.

```
struct VisorsGnc {
  VisorsGnc(VisorsGncOutput&);
  void in_bus_telemetry(…);
  void in_gps_message(…);
  void in_crosslink(…);
  void in_ground_command(…);
};

struct VisorsGncOutput {
  virtual void out_maneuver(…) = 0;
  virtual void out_observation(…) = 0;
  virtual void out_mission_mode(…) = 0;
  virtual void out_crosslink(…) = 0;
  virtual void out_telemetry(…) = 0;
};
```

**Figure 9.** Flight software event-driven inputs and outputs implemented in C++ as class member functions.

Explicit input and output functions allow the simulation to fully interface with the flight code and appropriately mock all sensors and actuators the flight software expects to interact with. In C++, the input/output functions take the form of class member functions, where the output functions are virtual in order to be implemented in either simulation or a flight computer host process (Figure 9).

In this work, the flight software is also explicitly designed to allow creating multiple instance of the flight software at once. Concretely, the flight software is intentionally written not to depend on global variables. Without this design constraint, running multiple instances of flight software in the single simulation process would not work.

*Scheduling*

Determinism is achieved by running all flight software processes inside a single operating system process, executed by a deterministic scheduler (Figure 10). Flight software is written to follow an event-driven message-passing style, so there is no need to preempt processes. Therefore, scheduling is fairly simple, achieved by adding invocations to flight software to the simulation event loop at the desired time (see Section 3). Times can be perturbed by pseudorandom noise to represent delayed messages from other spacecraft or scheduling noise, for example. At the time of writing, all flight software execution is serial (single-threaded). In the future, it would be possible to parallelize the execution of the simulation without violating the deterministic total ordering of events, in a style similar to parallel discrete-event simulation[64].

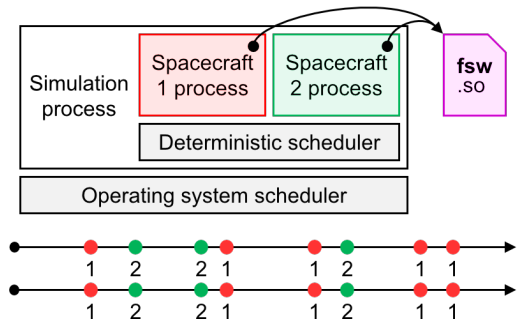

**Figure 10.** Flight software loaded as a shared library and executed by a deterministic simulation scheduler. Flight software execution order is identical from run to run.

Using a simulated scheduler within a single operating system process is crucial for providing (3) *deterministic total order*, (4) *faster-than-real-time execution*, and (5) *easy debugging*. It's enabled by compiling flight software to a shared library rather than an executable, which would require starting separate operating system processes. This design contrasts with the typical configuration for integrated testing of distributed software, in which multiple processes are started on either the same or separate computers and scheduled by the operating system (Figure 11).

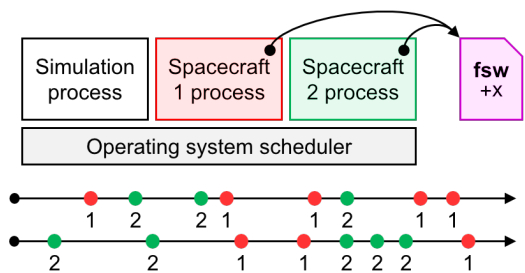

**Figure 11.** Flight software loaded as an executable and run by a nondeterministic operating system scheduler. Flight software execution order varies from run to run.

*Timed Execution*

A callback-based interface is used to allow flight software to execute code at specific times, to run timed events like maneuvers or timers. Like the input/output interface, the goals of the timed events interface are 1) to allow the simulation to intercept requests and substitute its own logic for scheduling flight software events in simulation, and 2) to be easy to implement correctly in the eventual host process on the real flight computer. To this end, the interface designates a special "tick" input to the flight software, which is a notification of the current time. Similarly, it designates a special "tick request" output, which constitutes a request from the flight software to its environment to input a tick at the given later time. Thus, flight software can accomplish timed execution by outputting a tick request for the desired time and, when the requested tick is later received, taking the desired action. Timed execution supports implementing the flight software in an event-driven style, enhancing capability (4) *run faster than real-time*.

In order to keep the external implementation simple, the tick interface assumes that the environment (either simulation or flight computer host) can only remember a single next tick. The requested tick output can be thought of as setting this single value. If flight software wants to schedule multiple timed events in the future, it should request a tick for only the first one, and after receiving it, request a tick for the second one, then the third, etc. The added flight software complexity to do this is marginal and even desirable, since it simplifies the implementation of the eventual host process.

*Memory Management*

In simulation, each flight software process is allocated its own memory allocator to track memory use. In order to allow the flight software to continue to use the familiar global `malloc`/`free` functions without interfering with each other, a `current_heap` pointer global variable is defined to allow changing which allocator is used at a given time. The `malloc`/`free`/`realloc`/`calloc` functions are substituted in the flight software via interposition[65] with versions that dispatch through the global variable. The `current_heap` is set before and cleared after invoking any code belonging to a flight software process (Figure 12). This is done via a C++ constructor/destructor pair and the RAII pattern[66] to ensure the pointer is always restored after flight software execution so as not to interfere with memory allocation in the simulation code, even in case of exceptional control flow. Tracking and using multiple heaps at once is a capability especially valuable for the development of distributed space flight software, although it could also be used for multiple processes on a single spacecraft.

*Limitation: Processor Architecture*—One limitation of the interface virtualization design in this work is that it cannot virtualize flight software compiled for a different processor architecture from the development computer. For example, if a flight computer runs ARMv7[67], and the development computer runs x86-64[68], the flight software must be compiled to an ARMv7 shared library for the flight computer and x86-64 shared library for the development computer. This is usually easy to do if the flight software is written in portable C, however it means that the flight software is not binary-identical to the code that will ultimately run on the spacecraft, though it is source identical. In practice, the wide availability of standards-conformant C and C++ compilers means this does not have a large impact on development, although it can occasionally cause behavior differences between test and deployment executables in case of things like pointer size and memory alignment. A possible solution to this problem would be to use hardware emulation tools like QEMU[69] or AVRS[70], which allow executing code for a different processor architecture than the development machine at the cost of some performance.

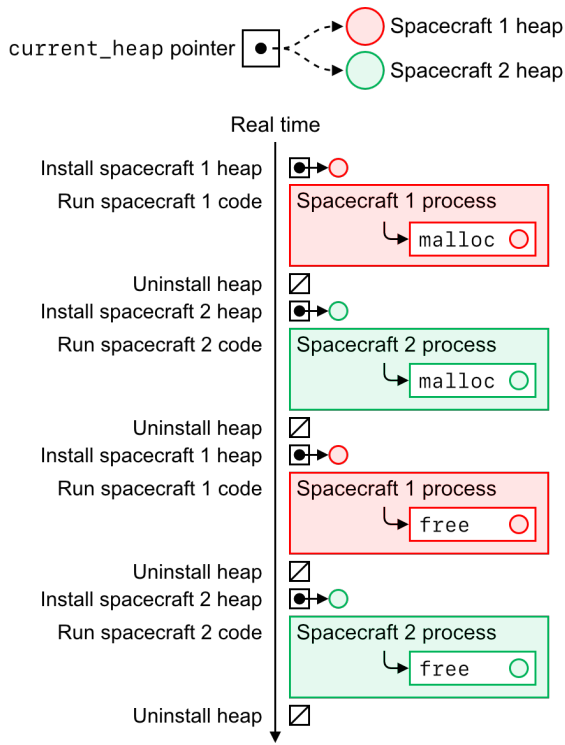

**Figure 12.** Global memory allocator functions are overridden to dispatch `malloc`/`free` calls from different flight software processes to different heaps.

## 5. ENVIRONMENT MODELS

The third main contribution of this work is the development and integration of a variety of high-fidelity models specifically for the distributed space flight software environment. Beyond high-fidelity force modeling, which is required for any space dynamics simulation, these models attempt to capture the operating environment of distributed space flight software as fully as possible. Of particular value to state-of-the-art distributed space systems are models for imperfect inter-satellite communication, software memory and runtime constraints, high-fidelity raw GPS measurements, and on-board cameras. Combined with event-driven simulation and effective virtual interfaces, these models allow representatively testing distributed space flight software regularly during iterative development, providing insight into software performance and defects.

*Orbit Dynamics*

Ground-truth dynamics modeling (Table 1) is based on the S³ astrodynamics library in the Stanford Space Rendezvous Laboratory[71], which has previously been validated against absolute and relative flight data from the PRISMA mission, accurate to meter- and centimeter-level respectively[72][73].

**Table 1. Ground-truth dynamics**

| Model | Implementation |
|---|---|
| Geopotential | GGM05S ($60 \times 60$)[74] |
| Atmosphere density | NRLMSISE-00[75] |
| Atmosphere drag | Cannon ball, $C_d = 2.2$ |
| Wind-relative velocity | Earth-fixed atmosphere |
| Solar radiation pressure | Analytical Sun ephemeris<br>Discrete conical shadow<br>Cannon ball, $C_r = 1.8$ |
| Third-body gravity | Analytical Sun/Moon ephemeris[76] |
| Integrator | RK4 |
| Step size | 1–10 s |
| State representation | Quasi-nonsingular elements |
| Earth reference frame | IAU 1980/1976[77] |

*GPS Receivers*

In order to enable development of precise real-time differential GPS navigation algorithms such as DiGiTaL, Novatel-brand GPS receivers for each spacecraft are simulated, and raw messages including range and carrier-phase are output in the Novatel binary message format. This supports developing flight software that can interface directly with Novatel GPS receivers in flight.

The simulation computes 31 operational GPS satellite orbits via closed-form perturbed orbit model and generates ranges by difference with each spacecraft's ground-truth position (Figure 13). Measurements are corrupted by noise based on performance reported by the manufacturer[78] (Table 2), and a precomputed GPS antenna gain pattern[79] dynamically selects the pseudorange and carrier phase noise standard deviation based on the elevation of each GPS satellite with respect to the antenna, given the spacecraft's current attitude (Figure 14). GPS signals are emitted at a regular cadence aligned with 1-second boundaries in GPS time. Line-of-sight visibility between each GPS satellite and simulated spacecraft are computed accounting for Earth occlusion and antenna attitude in order to provide the correct set of available ranges at each time step. Additionally, receiver-computed position-velocity-time solutions are modeled at each step by adding multivariate normally distributed noise to the receiver's true position and velocity.

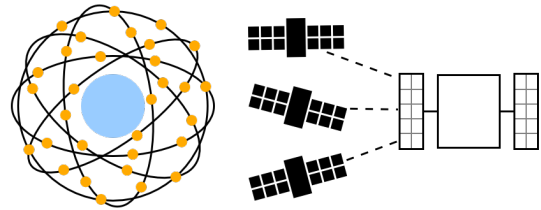

**Figure 13.** 31 operational GPS satellites are simulated, and individual range and carrier phase measurements are generated for each spacecraft.

**Table 2. GPS receiver/constellation model noise**

| Quantity | | Noise distribution |
|---|---|---|
| Pseudorange | $\rho_{pr}$ | $\mathcal{N}(0,\ 0.1437\ \text{m} \le \sigma \le 2.2769\ \text{m})$ |
| Carrier phase | $\rho_{cp}$ | $\mathcal{N}(0,\ 0.659\ \text{mm} \le \sigma \le 10.45\ \text{mm})$ |
| Position | $r$ | $\mathcal{N}(0,\ 1.5\ \text{m})$ |
| Velocity | $v$ | $\mathcal{N}(0,\ 30\ \text{mm/s})$ |
| Integer ambiguity | $N$ | $\mathcal{U}_{\mathbb{Z}}(-5,\ 5)$ |
| GPS vehicle RTN perturbation | $r_s$ | $\mathcal{N}(0,\ 1\ \text{m})$ |

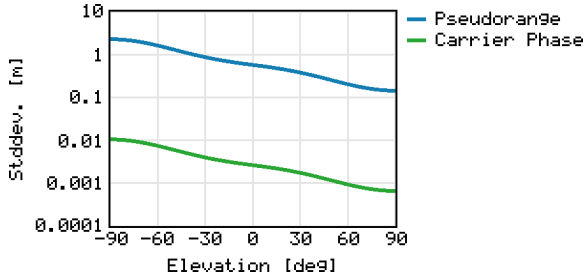

**Figure 14.** Elevation-dependent antenna gain pattern determines effective standard deviation for GPS pseudorange and carrier phase noise.

*Radio Communication*

Radio communication is a core function of any spacecraft, but it is especially important to consider when designing a distributed space system. Formation-flying may need more frequent communication with ground control than a single spacecraft would, and multiple spacecraft may rely on inter-satellite messaging for autonomy or navigation. For example, real-time differential GPS depends on the exchange of measurements between two spacecraft via crosslink.

Radio communication introduces two potential impacts to space flight software: 1) transmission delay, and 2) unreliable delivery. While simple, these effects can be difficult to design against without careful thought[80], and can exercise defective edge cases in flight software that are difficult to identify without involved testing and analysis[81,82].

Unreliable delivery can cause important information (such as commands) to be lost or, in attempt to avoid loss, duplicated. In this work, unreliable delivery is modeled using a per-link Markov chain between blackout-on and blackout-off states (Figure 15). When a message is sent via radio, the Markov chain is stepped probabilistically to either state according to defined transition probabilities from its current state. Messages are dropped while in the blackout-on state.

Transmission delay can cause messages to be delivered later than a desired deadline or out of order with respect to other related messages. In this work, transmission delay is modeled using a log-normal distribution, which has previously been used to model network delay[83,84]. The distribution parameters are adjustable; in this work we commonly choose $\mu$ and $\sigma$ such that transmission delays have $-3\sigma$ lower bound of 0.1 s and $+3\sigma$ upper bound of 10 s (Figure 16).

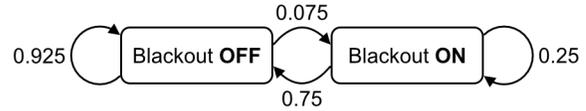

**Figure 15.** Stochastic blackout model for radio links.

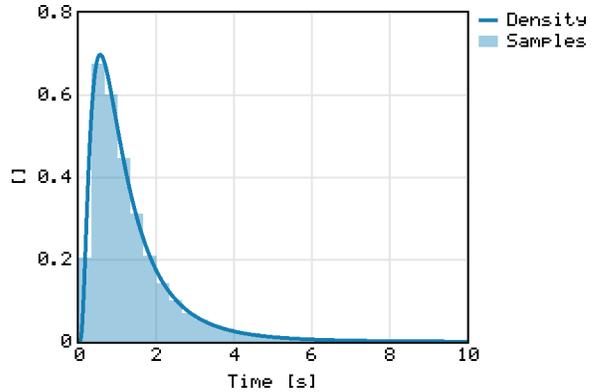

**Figure 16.** Probability density and 10,000-sample histogram of simulated transmission delay.

Combined, transmission delay and unreliable delivery can cause significant differences between sequences of sent and received messages (Figure 17). These differences can expose defects and deteriorate performance of distributed space flight software in unpredictable ways, and thus are crucial to model continuously during development.

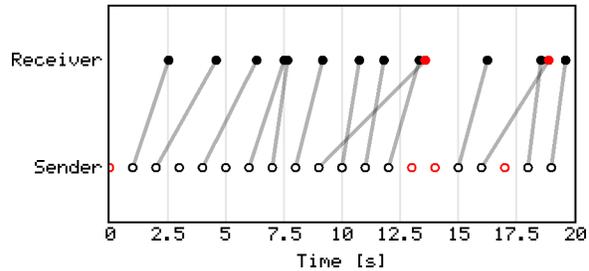

**Figure 17** Regularly sent vs. irregularly received messages with transmission delay and unreliable delivery; dropped and re-ordered messages are highlighted.

*Memory Allocation*

Flight software memory allocation is modeled in simulation using a simple heap allocator, which is exposed to the flight software by virtualizing the `malloc/free/calloc/realloc` functions. The simulated memory allocator is designed to conservatively model heap fragmentation relative to what might occur in a real flight software environment.

The simulated heap allocator arranges variable-length 8-byte aligned heap blocks contiguously in memory, abutted by 4-

byte headers and footers (Figure 18). Headers and footers store each block's size and whether it's currently allocated (Figure 19). Header and footer contents are identical. Headers and footers allow traversing blocks sequentially in memory for the sake of splitting blocks when allocating and coalescing adjacent blocks when freeing.

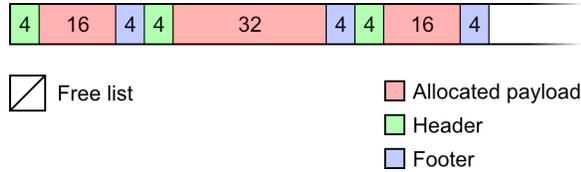

**Figure 18.** Memory layout of heap blocks.

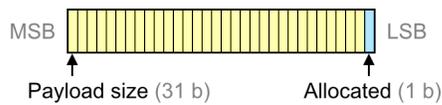

**Figure 19.** Contents of heap headers and footers.

Free blocks are tracked in an explicit doubly-linked free list; the first 8 payload bytes of each free block store the *next* and *previous* pointers. Figure 20 shows the free list after freeing the first and third block in Figure 18. Upon allocating (`malloc`), the free list is searched from head to tail (most to least recently freed) using a *first-fit* policy: choose the first block large enough to accommodate the request. If the block size is bigger than requested, the remaining space is split into a new block and returned to the free list. If the free list has no suitable block, a new block is created at the end of the heap.

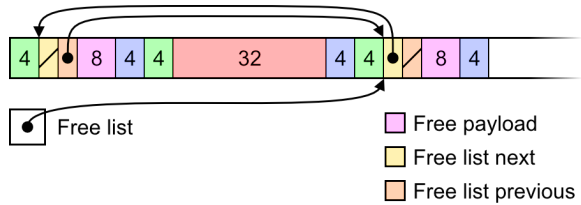

**Figure 20.** Free blocks stored in a doubly-linked free list.

When freeing (`free`), adjacent blocks are checked using headers and footers, and if they're free, the blocks are coalesced into a single larger free block (Figure 21). The resulting block is added to the front of the free list. The functions `calloc` and `realloc` are also virtualized. `realloc` also checks the following block's header and coalesces if possible for in-place reallocation.

This allocator design was chosen to conservatively model heap fragmentation that might occur in a real flight software environment. By using a single free list and first-fit policy, the simulated heap is intentionally more prone to fragmentation than a more sophisticated allocator while remaining reasonably efficient in throughput and utilization.

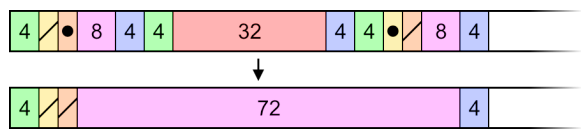

**Figure 21.** On `free`, adjacent free blocks are coalesced.

*On-Board Cameras*

Finally, of particular interest for next-generation computer vision applications in space are models for vision-based sensors. Far-range angles-only navigation using ARTMS[85] has recently been demonstrated successfully in orbit for the first time by the StarFOX experiment on-board the 2023 Starling technology demonstration mission[86], and new algorithms in the RPO Kit flight software package combine optical navigation algorithms from ARTMS and SPN[87][88] to perform new hybrid near- and far-range optical navigation for rendezvous and proximity operations.

The simulation environment in this work includes an OpenGL-based real-time rendering pipeline for generating imperfect images of the space environment that would be captured by cameras on board the spacecraft. The model can render the star field (with sub-pixel accuracy), Earth, and near-range target spacecraft based on the current orbit, attitude, and camera parameters of the observer (Figure 22). This work builds on the synthetic image generation capabilities of [89] and [90], and is the first instance in this line of work of running synthetic image generation in closed-loop with flight software that can actively influence the pose of the observer.

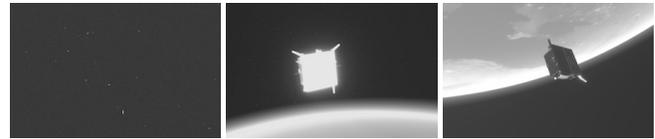

**Figure 22.** Synthetic images of far- and near-range targets generated using an OpenGL pipeline.

# 6. CASE STUDY RESULTS

Event-driven simulation was used for iterative development of two state-of-the-art GNC flight software packages: VISORS GNC and RPO Kit, developed by the Stanford Space Rendezvous Laboratory. VISORS GNC development began in January, 2022, and RPO Kit development began in June, 2023; development for both projects is ongoing as of October, 2024. During that time, the VISORS flight software had 579 revisions, and the RPO Kit flight software had 154 revisions. Table 3 compares the development period, number of revisions, and 50-/75-/90-th percentiles of modified lines of code per revision for both projects. Both projects were developed by small teams; on average VISORS GNC had 3–4 active contributors and RPO Kit had 2–3 in a given month. Hybrid event-driven simulation, software interface virtualization, and high-fidelity environment models were co-developed over time alongside the flight software (not included in revision statistics), and successfully enabled an iterative development process while detecting defects and characterizing software performance and reliability.

*Developer Workflow*—Most of the time, developers for VISORS GNC and RPO Kit used a local iterative development approach when revising software, in which they made code changes, ran one or more simulation scenarios on their development computer to evaluate the result of their change, and possibly made additional changes until they were happy with the simulation result. Most of the time, this workflow was sufficient to produce correct code within a single revision. Occasionally, a change introduced a defect that only very rarely caused detectable faults. These defects might not be detected in simulation until much later in development, possibly after many months. However, because simulations were deterministic, whenever a particular simulation case was discovered that detected a rare fault, the fault could easily be reproduced in order to resolve the defect. Hardware-in-the-loop testing was performed less frequently, and used to validate and improve simulated noise models.

**Table 3. Flight software revisions**

|  | VISORS GNC |  | RPO Kit |  |
|---|---|---|---|---|
| **Duration** | 33 mo |  | 16 mo |  |
| **Revisions** | 579 |  | 154 |  |
| **# Lines** | Added | Removed | Added | Removed |
| **P50** | 30 | 32 | 10 | 15 |
| **P75** | 111 | 104 | 38 | 54 |
| **P90** | 283 | 310 | 205 | 246 |

*Navigation and Control Performance*

While many of the contributions in this work are focused on software execution and distributed system modeling, the resulting high-fidelity simulation framework is still targeted at development of distributed navigation and control flight software, and it supports analysis of typical performance metrics like navigation error and control accuracy. Crucially, these metrics can be collected quickly and reliably from multiple interacting flight-ready software processes while incorporating variation in data availability, timing, memory allocation, and the space environment.

For example, Figure 23 shows navigation accuracy of the VISORS GNC flight software as of September, 2024, which demonstrates real-time differential GPS using integer ambiguity resolution to achieve relative navigation errors less than 1 cm using only GPS signals. This navigation accuracy can only be achieved using communication between the two spacecraft over a noisy inter-satellite link, and its performance depends on many factors, including successful delivery of packets. Degraded accuracy and 1-$\sigma$ error confidence can be seen around the 0.4-orbit mark, due to a combination of dropped crosslink packets and loss of visible GPS satellites due to slewing of the two spacecraft.

For another example, flight software simulation was used regularly during VISORS GNC development to verify the propulsion budget. The VISORS mission requires that the two-spacecraft formation reconfigure between different relative orbits. GNC is responsible for planning transfer trajectories for these reconfigurations, and they constitute a significant part of the $\Delta v$ budget for the mission. Figure 24 shows the result of 100 Monte Carlo simulations to assess the $\Delta v$ used for a single transfer from standby to science formation alignment. Each simulation uses slightly different initial conditions and model parameters, sampled from a user-defined distribution. Outside of this work, this kind of $\Delta v$ verification might be limited to the preliminary design phase of the mission or performed with simplified models of the flight software; here, an up-to-date $\Delta v$ analysis of the flight-ready software is always available and easy to obtain.

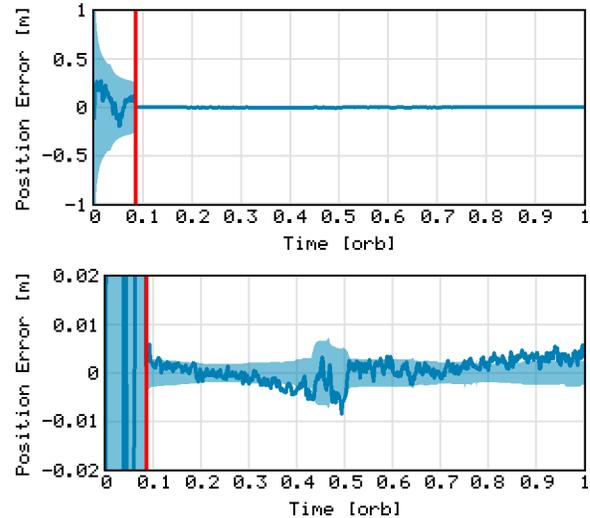

**Figure 23.** VISORS relative navigation error and 1-$\sigma$ confidence over first orbit after initialization, estimated by spacecraft 0. Top/bottom plots are identical but for *y*-axis zoom. High-precision navigation with integer ambiguity resolution begins after ~8 min (red line).

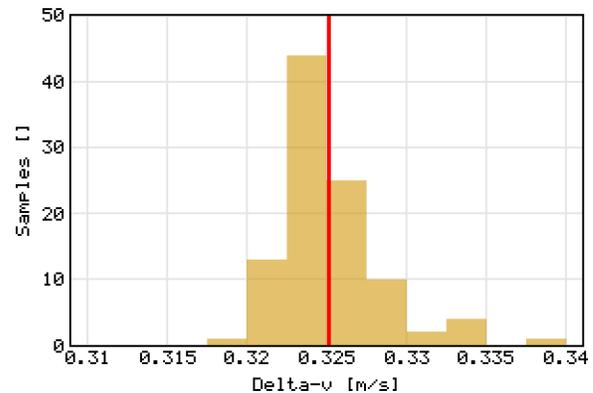

**Figure 24.** Histogram of $\Delta v$ required for a single transfer trajectory planned by VISORS GNC across 100 Monte Carlo simulations. Mean ≈ 0.325 m/s (red line).

*Simulation Speed*

By using hybrid event-driven simulation, idle time between flight software executions can be skipped entirely, so simulations execute quickly even while accurately modeling the flow of time and synchronization between distributed software processes. The 100 Monte Carlo simulations in Figure 24 were run on an Apple M3 Max processor and took an average of 9.53 seconds real-time to simulate an average of 20.05 hours of simulation time—a speedup of about 7,500x. Running quickly while fully capturing important details like network delays, dropped packets, and spacecraft orbit and attitude is a crucial part of enabling an iterative software development flow, since changes to flight software can be evaluated as quickly as they can be implemented.

*Determinism*

All noise models in the simulation environment use pseudo-random noise derived from a specified seed, and distributed flight software processes are executed by the simulation in a fully-determined total order, so the entire evolution and final state of a simulation is deterministic. This affords the unique ability to run multiple interacting flight software processes with high-fidelity timing, communication delays, and noisy sensors and actuators, with full reproducibility. If an error occurs in a process several minutes into a simulation (which may represent several days of simulated time), the simulation can easily be restarted and run with additional instrumentation to inspect the state and identify the source of the error. This is invaluable for the development of distributed space systems, and allows aggressively resolving defects in complex interactions between distributed flight software processes even if they present themselves only rarely.

Determinism was prized highly during development of VISORS GNC and RPO Kit, and simulation results were continuously monitored using a few complimentary techniques to ensure determinism was never violated. First, at the end of each simulation, a 32-bit integer was generated using the global pseudorandom number generator and recorded as a fingerprint. Matching fingerprints from repeated simulations indicated that the same pattern of random values, and likely pattern of software execution, was followed[91]. Second, the SHA-256 hash of analysis output was periodically checked after repeated simulations to ensure that a variety of high-level metrics were identical. Third, individual simulation outputs were inspected manually. Table 4 shows examples of these determinism checks after three runs of the same simulation. On one occasion, in December, 2023, determinism was found to be violated, which was studied with high priority. The cause was that an external dependency was assumed to be thread-safe because it exposed a pure functional interface, but it in fact used unprotected global variables to pass data between functions internally (reinforcing the recommendation to avoid using global variables).

**Table 4. Example determinism checks for three runs**

| Run | Fingerprint | Analysis hash | Mean position error |
|---|---|---|---|
| 1 | e7174029 | 87c6affe… | 14.592582331956 |
| 2 | e7174029 | 87c6affe… | 14.592582331956 |
| 3 | e7174029 | 87c6affe… | 14.592582331956 |

*Detecting Flight Software Defects*

The ability to run deterministic closed-loop simulations of multiple interacting flight software processes with dynamics, sensors, and actuators proved invaluable for detecting and identifying defects during VISORS GNC and RPO Kit. development. Most of the time, defects were detected and resolved while implementing changes within a single software revision. However, sometimes a defect presented itself only in rare cases and required later work dedicated to identifying and resolving it. Here, a few defects are described.

*Dropped Crosslink Messages*

Synchronizing information between separate spacecraft is a crucial capability for autonomous distributed space systems. As with distributed computer systems on Earth, communication between spacecraft must be robust to missing, late, and duplicate messages, and implementing communication protocols correctly is challenging. Event-driven simulation aided in detecting defects in VISORS GNC crosslink communication via its ability to model small variations in time and order of events in a hybrid discrete-continuous system.

Early in VISORS GNC development, a rudimentary design was implemented for synchronizing its state machine between the two spacecraft, in which event transitions were sent via crosslink from one spacecraft (designated active) to the other (designated passive). These event transitions were sent whenever they occurred in the active spacecraft, without retransmitting or re-ordering dropped or late messages.

Using this design in the final flight software would have been incorrect, since individual events could be dropped or duplicated on the receiving spacecraft and lead to invalid transitions. For example, a simplified model of part of the state machine is shown in Figure 25; the machine can switch between science mode and taking an observation via begin-observation and end-observation events. It's invalid to begin an observation if already in observing mode or end and observation if not currently observing.

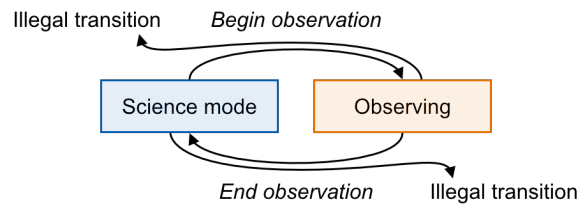

**Figure 25.** Simplified model of a VISORS GNC state machine that switches between science mode and observing in response to begin-observation and end-observation events.

Event-driven simulation and crosslink communication modeling was used to detect these faults by randomly dropping and delaying crosslink messages. This successfully detected the invalid transitions mentioned above by causing program crashes. An example execution from such a simulation is shown in Figure 26. Detecting such faults in simulation ultimately allowed iteratively developing an alternative synchronization design in July, 2024, that communicates transitions preemptively and retransmits if they are dropped.

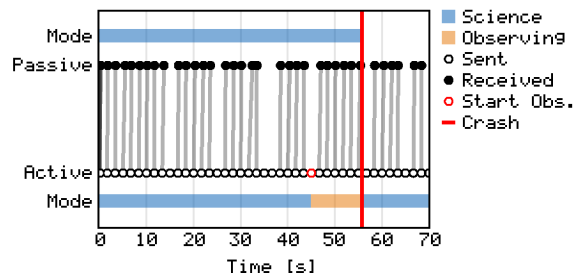

**Figure 26.** Program crash caused by dropped state machine events between active and passive spacecraft leading to an invalid transition.

*Reordered Crosslink Messages*

In addition to state machine synchronization, crosslink unreliability also allowed detecting bugs in the VISORS GNC and RPO Kit navigation software. The close-proximity formation-flying capabilities of both systems rely on precise position knowledge from real-time differential GPS, which requires exchanging GPS measurements via crosslink. The navigation component maintains a queue of received GPS messages from the opposite spacecraft, which mistakenly assumed measurements arrived from the remote spacecraft in sorted order despite the VISORS GNC interface specifying robustness to re-ordered messages. This resulted in a bug, which was identified via simulation (Figure 27). The defect was resolved for the short term by simply ignoring the late measurement; in the future, a more resilient way of still making use of late data may be developed.

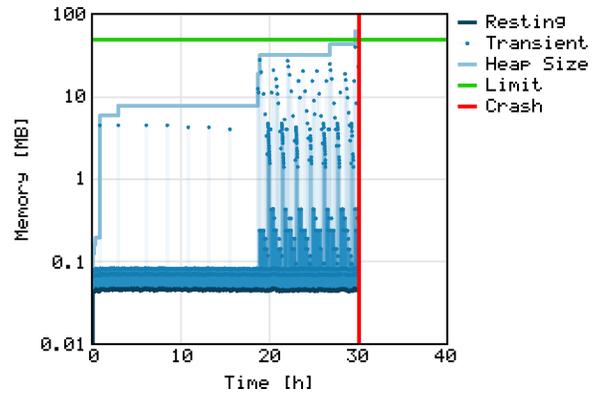

**Figure 28.** VISORS GNC memory use and program crash due to exhaustion of 50-MB heap limit.

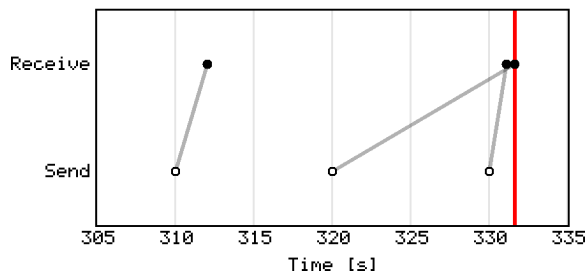

**Figure 27.** Long-delayed earlier crosslink message received after short-delayed later crosslink message, causing a software crash (red line) in the navigation queue.

*Memory Exhaustion*

Space flight computers often have less memory than terrestrial computers, and space flight software may be long-running, so memory leaks and fragmentation are important to protect against. While some advise to avoid dynamic memory allocation altogether[92], in practice using limited dynamic allocation usually allows for simpler and more memory-efficient programs[93]. In these cases, safety is often achieved by budgeting a predetermined amount of memory to different components of a system. For example, a requirement for VISORS GNC is to use a maximum of 50 MB of dynamic memory. During development, high-fidelity simulation was used to detect excessive memory use by simulating dynamic memory allocation on each spacecraft while executing the flight software and intentionally crashing the program if it ever exceeded the 50-MB limit.

As of April, 2024, VISORS GNC occasionally exceeded the 50-MB dynamic memory limit, causing a program crash. Figure 28 shows resting and transient memory and total heap size on the active spacecraft over a 50-hour science campaign. Resting memory refers to allocations that persist in between flight software invocations; transient memory refers to memory allocated while running the flight software but freed before the flight software finishes. As seen in the figure, the majority of memory used by GNC is transient.

The primary source of memory exhaustion was identified using debugging tools on the program crash site. One function of VISORS GNC is to split few large impulsive maneuvers into many small impulses spread out over time to make them realizable by the impulse-limited propulsion system[94]. To split maneuvers, GNC relies first on numerical optimization and second on an analytic fallback procedure. Numerical optimization uses ECOS[95], an open-source second-order cone program solver for solving problems of the form

$$\begin{aligned} \min \quad & c^\top x \\ \text{subject to} \quad & A\,x = b \\ & G\,x \leq_{\mathcal{K}} h, \end{aligned}$$

with matrices $A$ and $G$, vectors $h$, $b$, and $c$, and decision variable $x$, where $\mathcal{K}$ is a cone. In practice, $G$ is very large for the maneuver-splitting problem and used the majority of memory within the GNC system. This caused a software crash if a large maneuver-splitting problem was attempted.

In this case, the defect was resolved with a sparse matrix optimization. The matrix is known a-priori to hold zero elements off the diagonal, so most of the memory allocated to the matrix does not store significant information. ECOS natively supports a sparse input format in which only non-zero elements are specified (Figure 29). Switching to this format produced a significant memory reduction and factor of safety from exceeding the required memory limit during the same scenario (Figure 30).

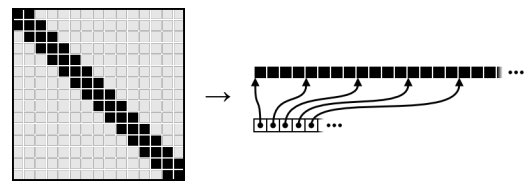

**Figure 29.** Dense mostly-zero matrix replaced with sparse matrix representation as input to second-order cone program in order to reduce memory use.

*Fragmentation*—Although memory consumption was reduced, a net increase in the total heap size can be observed from the start to the end of the science campaign. While this is expected given the transient memory needs of the science campaign, it is important to determine whether this increase is stable or would grow if another science campaign were performed. During VISORS GNC development, long-duration Monte Carlo testing was used to successfully verify that fragmentation did not cause memory exhaustion over the course of six months of continuous operation.

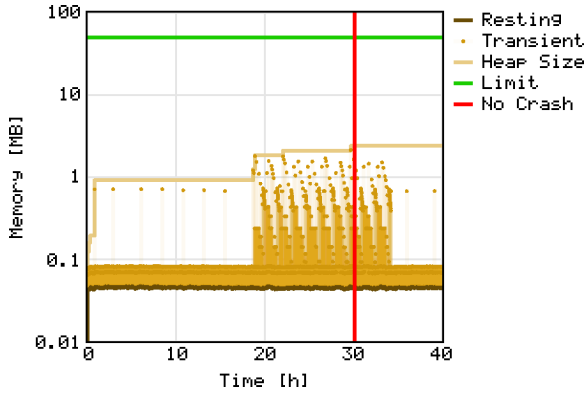

**Figure 30.** Resolved VISORS GNC memory use with sparse matrices during science campaign, with no crash and ~20x safety factor against memory exhaustion.

Additional defects detected and resolved via extended simulation testing but not described in detail here include:

- (RPO Kit navigation) Rare failure to Cholesky-decompose positive-semidefinite covariance matrix; caused by integrating two flight software components with previously incompatible covariance matrix conventions; resolved by changing the ordering of states in the covariance matrix.
- (VISORS closed-form control) Rare failure to plan maneuvers due to unhandled `atan2` angle-wrapping inside Newton-Raphson iteration within a closed-form solver; resolved by checking for angle wrapping after `atan2`.

On the whole, detecting and resolving software defects in simulation, including those due to unreliable communication and memory limitations, enabled making software changes with greater confidence than if the software had required real-time nondeterministic testing. The capabilities to simulate long stretches of operation quickly and to reliably reproduce rare failures once they were discovered were especially useful, since they completely eliminated the common developer experience of bugs that are known to exist but hard to observe[96].

*Processor-in-the-Loop Testing*

While high-fidelity simulation testing is a critical enabler of rapid, iterative development, hardware-in-the-loop testing is necessary to validate the simulation. Hardware testing has the potential to identify defects not detected in simulation, in which case the simulation must be improved.

During VISORS GNC development, processor-in-the-loop testing was used to ensure proper communication between distributed GNC processes and proper functioning when run on a 32-bit ARM instruction set architecture, which is representative of the VISORS flight computer. The flight software shared library ran inside a host process on each of two BCM2835 processors (ARMv6 architecture)[97], while the simulation ran in closed-loop on a consumer laptop (Figure 31). Crosslink communication between the flight software used the laptop and simulation environment as a relay. This configuration was first attempted in August, 2023, after about 18 months of VISORS GNC development.

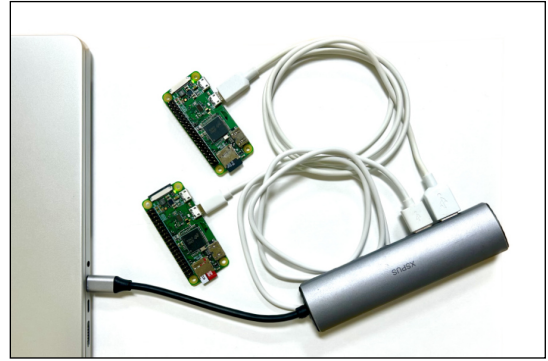

**Figure 31.** Running two-spacecraft distributed flight software on BCM2835 processors (ARMv6 architecture) via the Raspberry Pi Zero W single-board computer.

After implementing a host process to wrap the flight software on the external processor, a simulation of the VISORS mission was successfully run within the same day, and performance metrics collected. Table 5 compares selected results from processor-in-the-loop testing with simulations run entirely in software on the development computer. Metrics include successful observation count as a bottom-line control objective and mean and max runtimes for four flight software algorithms used for navigation: group and phase ionospheric correction (GRAPHIC) measurement update[98], single-difference carrier phase (SDCP) update, filter time update, and integer ambiguity resolution (IAR).

**Table 5. Processor-in-the-loop test results**

| Metric | ARM (BCM2835) | Desktop (M3 Max) |
|---|---|---|
| # good observations | 7/10* | 10/10* |

| | Runtime | | | |
|---|---|---|---|---|
| Algorithm | Mean | Max | Mean | Max |
| GRAPHIC update | 23.5 ms | 156.0 ms | 344 µs | 1,121 µs |
| SDCP update | 16.2 ms | 100.9 ms | 246 µs | 804 µs |
| Time update | 2.2 ms | 6.9 ms | 26 µs | 116 µs |
| IAR | 0.5 ms | 4.0 ms | 2 µs | 55 µs |

\* Successful observation count is a noisy metric that varies across simulations; the difference shown here is merely for example and not statistically significant.

Performing successful hardware-in-the-loop testing on the first attempt, within a single day and with little prior preparation, reflects the power of event-driven simulation to enable developing capable distributed space flight software quickly and conveniently, without relying on real-time or hardware-in-the-loop testing to validate every change.

*Interactive Debugging*

Since all flight software processes run in a single system thread, interactive debugging was commonly used to pause and inspect flight software state during failures (Figure 32).

Due to liberal use of runtime assertions in the flight software, identifying defects usually consisted of running a failing simulation in a debugger, seeing where it failed, and determining why the code was incorrect. The interface virtualization design used in this work has the benefit that the internal state of a flight software process can be inspected at any time. Team members used GDB, LLDB, or the Windows debugger depending on their development platform.

*Integrated CPU Profiling*

Similar to interactive debugging, running all flight software, ground software, and simulation models in a single operating system thread enabled using off-the-shelf CPU profilers to understand and improve the time efficiency of flight algorithms and the system as a whole (Figure 33). This is especially useful for space applications, since space-grade microprocessors are usually much slower than consumer-grade processors. Profiling and tuning the simulation as a whole also helps maintain a fast, iterative development workflow.

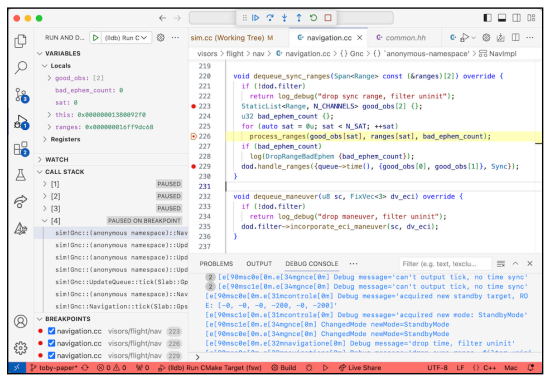

**Figure 32.** Off-the-shelf interactive debuggers were used to pause and inspect flight software on separate spacecraft at the same time.

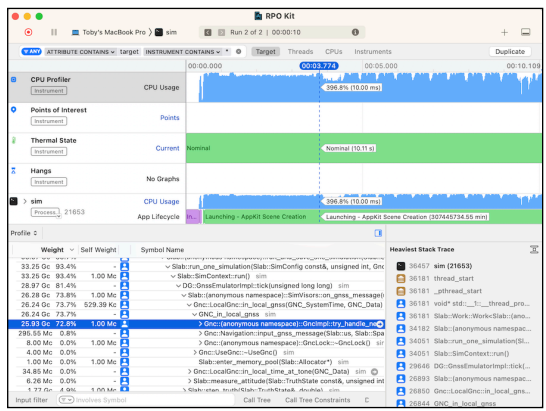

**Figure 33.** Apple Instruments was used on MacOS to profile and speed up both flight and simulation code.

# 7. CONCLUSION

A novel simulation environment based on hybrid discrete-continuous simulation was developed and used to enable rapid iterative development of two flight software projects for distributed space systems, VISORS GNC and RPO Kit. The simulation implements a virtual flight computer interface to allow executing unmodified compiled flight software, and includes a variety of models for the distributed flight software environment in order to assess robustness of flight software to the challenges of distributed spaceflight—most notably inter-spacecraft communication.

The simulation environment's unique combination of deterministic, faster-than-realtime execution and complex, noisy environment modeling was found to be especially powerful for testing distributed space flight software. Compared to typical navigation- and control-focused astrodynamics simulations as might be implemented in MATLAB/Simulink, the simulation environment in this work was found to be comparatively more powerful at detecting implementation-related defects like memory exhaustion, fragile communication protocols, and rare edge cases, without sacrificing the ability analyze key navigation and control performance metrics. On the other hand, compared to real-time software- and hardware-in-the-loop testing, the simulation environment in this work was found to be comparatively easier to use and more efficient for developers thanks to offering determinism and transparent debugging, while still capturing the benefits of real-time hardware-in-the-loop testing for rigorously exercising distributed software interactions. Thus, event-driven virtualized software simulation can be seen to offer a compelling combination of the advantages of both pure dynamics simulation and real-time integration testing for distributed space flight software developers.

Future work includes making the simulation environment user-friendly so it can be used by a non-expert, continued characterization and refinement of the built-in environment models' fidelity against hardware testbeds and past and future flight data, application to additional flight software projects, and dissemination of the tools and flight software as source code. Methods for parallel/multi-thread execution of a single simulation without losing determinism should be explored in order to reduce the time it takes to run simulations and accelerate iterative development.

An overarching limitation of the simulation environment in its current form is that it imposes requirements on the design of the flight software. For example, it requires that flight software be compiled to a shared library, not use any global state, and not start any operating system threads. In general, it requires flight software to be quite cooperative in order to interface with the simulation as a virtual environment. This is not a problem for new flight software projects, but it limits the ease with which this tool can be retrofitted to simulating existing standalone flight software executables. Relatedly, the simulation cannot execute flight software for different instruction set architectures than the development computer, and may require compiling the flight software specifically for simulation if the flight computer uses a different instruction set. This arguably violates the philosophy of running "unmodified" flight software. A possible solution to this problem would be to use hardware emulation tools like QEMU or AVRS. Overall, methods to reduce the constraints placed on the flight software should be explored.

A great deal of additional validation and refinement against real hardware and flight data should be performed on many of the models included in the simulation environment. In particular, synthetic image generation capabilities have been prototyped and shown to work with optical navigation algorithms, but generated images have not been rigorously compared to flight data. Memory allocation and fragmentation modeling for flight software should be compared to system `malloc`/`free` performance in a selection of real flight computer environments. Realistic sensor and actuation models, such as for propulsion, should continue to be developed to support additional spacecraft configurations. Finally, simulation capabilities focused on attitude determination and control should be developed and integrated, since thus far the focus of this work has been primarily on translational/orbit navigation and control flight software.

## ACKNOWLEDGEMENTS

This work was completed in association with the VISORS mission and RPO kit, supported by NSF award no. 1936663 and SpaceWERX Orbital Prime–Direct to Phase II contract no. FA864923P0560, respectively. The authors would also like to acknowledge Samuel Low for contributions to the GPS simulator noise model, and colleagues at the Stanford Space Rendezvous Lab for collaborative input.

## BIOGRAPHY

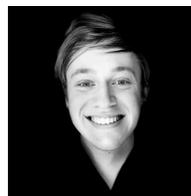

***Toby Bell*** *is a PhD student in the Space Rendezvous Laboratory at Stanford University, advised by Simone D'Amico. He researches space flight software simulation and optimization for distributed space systems. He received his BS and MS in computer science from Stanford University and previously was a flight software engineer at Astranis Space Technologies and Starlink process engineer at SpaceX. His hobbies include choral singing, social dancing, and building compilers.*

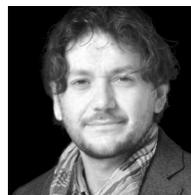

***Simone D'Amico*** *is Associate Professor of Aeronautics and Astronautics (AA), W.M. Keck Faculty Scholar in the School of Engineering, and Professor of Geophysics (by Courtesy). He is the Founding Director of the Stanford Space Rendezvous Laboratory, Co-Director of the Center for AEroSpace Autonomy Research (CAESAR), and Director of the Undergraduate Program in Aerospace Engineering at Stanford. He has 20+ years of experience in research and development of autonomous spacecraft and distributed space systems. He developed the distributed guidance, navigation, and control (GNC) system of several for-* *mation-flying and rendezvous missions and is currently the institutional PI of four autonomous satellite swarms funded by NASA (STARLING, STARI) and NSF (VISORS, SWARM-EX) with one of them operational in orbit right now (Starling). Besides academia, Dr. D'Amico is on the advisory board of four space start-ups focusing on distributed space systems for future applications in SAR remote sensing, orbital lifetime prolongation, and space-based solar power. He was the recipient of several awards, most recently the 2024 NASA Ames Honor Award for the Starling mission, Best Paper Awards at IAF (2022), IEEE (2021), AIAA (2021), AAS (2019) conferences, and the M. Barry Carlton Award by IEEE (2020). He received the B.S. and M.S. degrees from Politecnico di Milano (2003) and the Ph.D. degree from Delft University of Technology (2010).*